\documentclass[aps,eqsecnum,nofootinbib,superscriptaddress,showpacs]{revtex4-2}
\def\be{\begin{eqnarray}}
\def\ee{\end{eqnarray}}
\usepackage[dvipdfmx]{graphicx}
\usepackage{amsfonts,bm}
\usepackage{amsmath,amsfonts,amssymb,bm}
\usepackage[dvips]{color}
\def\p{\partial}

\def\ben{\begin{equation}}
\def\een{\end{equation}}
\def\bena{\begin{eqnarray}}
\def\eena{\end{eqnarray}}

\def\mS{{\mathbb S}}
\def\mV{{\mathbb V}}

\begin{document}
\title{The first law of entanglement entropy in AdS black hole backgrounds}

\author{Akihiro {\sc Ishibashi}}\email[]{akihiro@phys.kindai.ac.jp}
\affiliation{%
{\it Department of Physics and Research Institute for Science and Technology, Kindai University, Higashi-Osaka 577-8502, JAPAN
}}

\author{Kengo {\sc Maeda}}\email[]{maeda302@sic.shibaura-it.ac.jp}
\affiliation{%
{\it Faculty of Engineering,
Shibaura Institute of Technology, Saitama 330-8570, JAPAN}}

\begin{abstract}
The first law for entanglement entropy in CFT in an odd-dimensional asymptotically AdS black hole is studied by using the AdS/CFT duality. 
The entropy of CFT considered here is due to the entanglement between two subsystems separated by the horizon of the AdS black hole, which itself is realized as the conformal boundary of a black droplet in even-dimensional global AdS bulk spacetime. 
In $(2+1)$-dimensional CFT, the first law is shown to be always satisfied by analyzing a class of metric perturbations of the exact solution of a $4$-dimensional black droplet. 
In $(4+1)$-dimensions, the first law for CFT is shown to hold under the Neumann boundary condition at a certain bulk hypersurface anchored to the conformal boundary of the boundary AdS black hole. From the boundary view point, this Neumann condition yields there being no energy flux across the boundary of the boundary AdS black hole.  
Furthermore, the asymptotic geometry of a $6$-dimensional small AdS black droplet is constructed as the gravity dual of our $(4+1)$-dimensional CFT, which 
exhibits a negative energy near the spatial infinity, as expected from vacuum polarization. 
\end{abstract}
\maketitle

\section{Introduction}
The entanglement entropy of a quantum field in black hole backgrounds has attracted much attention as the key concept 
toward understanding the origin of the Bekenstein-Hawking entropy~(for review, see e.g.,~\cite{Solodukhin2011}). 
Although it is in general a Herculean task to calculate an entropy in a curved background, it is tractable to compute the entropy for 
strongly coupled conformal fields in the holographic setting, or the anti-de Sitter (AdS) conformal field theory (CFT) correspondence, 
by using the Ryu-Takayanagi formula~\cite{Ryu-Takayanagi2006}. In this formula, the entropy calculation for a subsystem is 
reduced to a much simpler task of calculating the area of a minimal or extremal surface anchored 
to the boundary of the asymptotically AdS gravity dual. 

In the framework of the AdS/CFT duality, the first law of the vacuum entanglement entropy was shown for 
any ball-shaped subregions when the AdS boundary is flat Minkowski spacetime~\cite{BCHM2013, WKZV2013}. 
Conversely, the bulk linearized Einstein equations around the pure AdS spacetime can be derived when the first law is satisfied for the subregions in the boundary theory~\cite{FGHMR2014}, being consistent with the basic idea~\cite{Jacobson95}. In some sense, the first law of the entanglement 
entropy could be a guiding principle for a consistent formulation of quantum gravity, just like the first law of black hole thermodynamics. 
Motivated by this, we holographically explore the first law of the entanglement entropy for two subsystems of CFT separated by a black hole horizon in asymptotically AdS spacetime.   

In the holographic proof of the first law \cite{FGHMR2014}, the Noether charge formula~\cite{IyerWald1994} plays an essential role.  
This is because the entanglement entropy for any ball-shaped spatial region 
in flat AdS boundary corresponds, through a conformal transformation, to the horizon area of a zero-mass hyperbolic black hole in the bulk. 
Therefore it is inferred that the Noether charge formula can also be applied to the derivation of the first law of the entanglement entropy of 
a strongly coupled CFT in black hole backgrounds. 
In the standard Noether charge derivation~\cite{IyerWald1994} of the first law of the black hole thermodynamics in $D$-dimensions, 
one first considers a $(D-1)$-dimensional spacelike hypersurface which has two $(D-2)$-dimensional boundaries, one at the black hole horizon (bifurcate surface) and the other at the spatial infinity, 
and then computes the Noether charges at these two boundaries. For asymptotically flat black holes, one can impose a natural boundary condition for physical fields at the spatial infinity to be consistent with the asymptotic flatness and black hole no-hair property, but for asymptotically AdS black holes, one needs to be more careful to consider possible boundary conditions at the AdS infinity.   
In the holographic setting, further care may be needed since one first extends the spacelike hypersurface 
into the $(D+1)$-dimensional bulk spacetime to have a $D$-dimensional spacelike hypersurface and, if the extended hypersurface 
ends at a bulk boundary (e.g., black hole horizon isolated from the boundary black hole), then one also has to 
take into account the additional contribution to the Noether charge from such a bulk boundary.  

In this paper, we examine the first law of the entanglement entropy of odd-dimensional conformal field theories~(CFT) in AdS black hole backgrounds by using the Noether charge formula in the holographic 
setting. Namely, we consider a $D$(odd)-dimensional AdS black hole 
which is realized as a part of the conformal boundary of the $(D+1)$-asymptotically AdS bulk spacetime. 
Then, as briefly mentioned above, we evaluate Noether charges at two $(D-2)$-boundaries (i.e., the boundary black hole horizon and the boundary 
AdS infinity) as well as those at $(D-1)$-bulk boundaries. 

For simplicity, we assume that the bulk geometry contains neither horizons in the IR isolated 
from the AdS boundary nor black funnels in the Hartle-Hawking state, which connect two or more boundary black holes~\cite{HubenyMarolfRangamani}. 
So, the only possible bulk solution dual to the boundary CFT in an AdS black hole background 
is the black droplet 
solution~\cite{HubenyMarolfRangamani}, in which the bulk horizon is connected to the single boundary black hole. 
The black droplet solutions with a boundary black hole in asymptotically flat spacetime were eagerly constructed with numerics, 
motivated by the investigation of the Hawking radiation in strongly coupled field 
theories~\cite{FLT2011,FT2013,FischettiSantos2013,SantosWay2014,Mefford2017,FischettiSantosWay2017}~(see also analytic solutions 
for lower-dimensional black droplets~\cite{HubenyMarolfRangamani} or 
perturbative analytic solutions~\cite{Haddad2012, IMM2017}).

\begin{figure}
 \begin{center}
  \includegraphics[width=80mm]{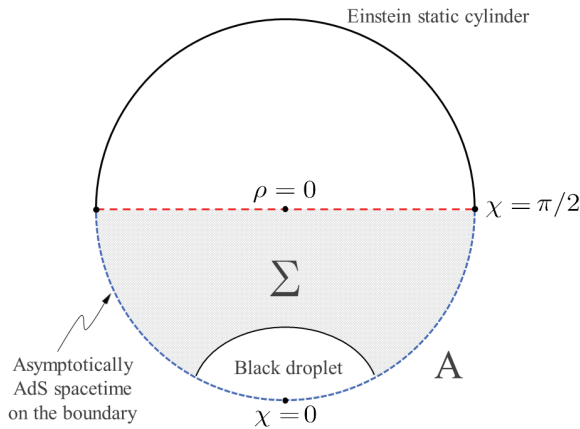}  
   \caption{A schematic diagram of a time-slice of our black droplet in asymptotically AdS bulk spacetime of $(D+1)$-dimension. 
  The black circle describes a $(D-1)$-dimensional sphere as a time-slice of the $D$-dimensional conformal boundary of the Einstein static cylinder or an asymptotically, globally AdS spacetime in $(D+1)$-dimension. 
 The conformal boundary of our black droplet itself is a $D$-dimensional asymptotically AdS black hole (illustrated in Fig.~\ref{fig2}), whose $(D-1)$-dimensional time-slice 
 (dashed blue half-circle) therefore covers only the half of the $(D-1)$-dimensional sphere. 
 Note that the boundary $D$-dimensional black hole has two asymptotic regions (corresponding to A and B of Fig. \ref{fig2}), but the lower half circle (dashed blue half-circle) shows only one of them, say the region A of Fig. \ref{fig2}. We take our hypersurface $\Sigma$ as the lower half of this time-slice (shaded region) enclosed by part of the AdS boundary (part of the dashed blue curve, which corresponds to the region A), black droplet horizon (black curve) and $\chi= \pi/2$ (red dotted line). The contribution to the Noether charge arises from part (red dotted line, $\chi=\pi/2$) of $\partial \Sigma$, 
as well as from the bulk horizon and the lower half of the AdS boundary. 
As we will discuss, we consider a reflection boundary condition 
at $\chi=\pi/2$ (red dotted line). 
}
\label{fig1}
\end{center}
\end{figure}

In Fig.~\ref{fig1}, we show an example of the black droplet solution in which the bulk horizon is connected to a boundary black hole in an asymptotically AdS spacetime.   
In $(2+1)$-dimensional field theory, we calculate the perturbation of the black droplet solution with BTZ black hole 
on the AdS boundary~\cite{HubenyMarolfRangamani}.~(For a brane model, the first law of the entanglement entropy 
was shown~\cite{EmparanFrassinoWay2020}.) 
It is shown that the first law of the entanglement entropy is always 
satisfied. In $(4+1)$-dimensional field theory, we show that the first law is satisfied for the Neumann boundary 
condition at the equatorial plane in Fig.~\ref{fig1}. In the field theory side, the condition corresponds to the 
no energy flux condition across the timelike null infinity. Under the boundary condition, we analytically construct the 
asymptotic geometry of such a black droplet solution from the perturbation of the global AdS spacetime. 
It is shown that negative energy appears in asymptotic region and it decays with $r^{-3}$ power in the radial coordinate $r$. 
This could be understood as a vacuum polarization effect. 
 
The rest of our paper is organized as follows. In section \ref{sec:2}, we briefly review the Noether charge formula~\cite{IyerWald1994} 
and explain our setup. In Sec.~\ref{sec:3}, we show that the first law of the entanglement entropy is satisfied in 
$(2+1)$-dimensional CFT by perturbing the black droplet solution. In Sec.~\ref{sec:4}, we derive the Noether charge on the equatorial plane 
and explore the boundary condition in which the first law is satisfied in $(4+1)$-dimensional CFT. We also show that 
the energy flux across the AdS boundary is zero when the first law is satisfied. In Sec.~\ref{sec:5}, a black droplet solution with a boundary 
AdS black hole is perturbatively constructed from the global AdS spacetime. Sec.~\ref{sec:6} is devoted to summary and discussions. 
In Appendix, we briefly discuss the Noether charge formula (\ref{first_law_form}) for general perturbations in $(5+1)$-dimensional bulk metric~(\ref{global_AdS}).

\section{Preliminaries: Holographic setting and Noether charge formula}
\label{sec:2}
\begin{figure}
 \begin{center}
  \includegraphics[width=80mm]{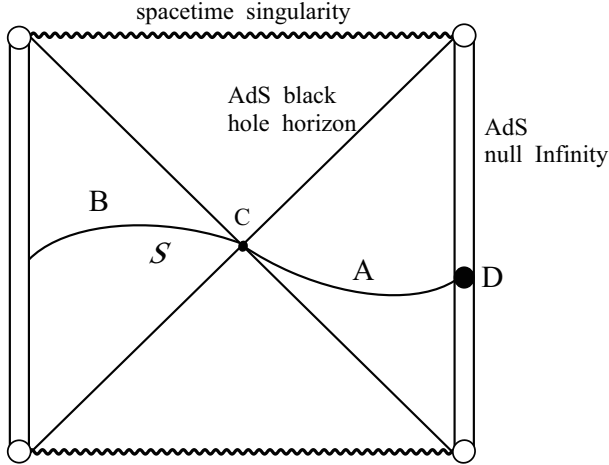}
  \caption{The Penrose diagram of an AdS black hole on the boundary theory. The Noether charge generically appears 
  on the bulk hypersurface anchored to ${\rm D}$.}
\label{fig2}
 \end{center}
\end{figure}
In this section we explain our setup and the strategy before exploring the first law of the entanglement entropy.    
Our holographic setup is the following: 
We consider a $(D+1)$-dimensional asymptotically (locally) anti-de Sitter~(AlAdS) spacetime $( {\cal M}, g_{MN})$ which satisfies 
the vacuum Einstein equations 
and whose conformal boundary $(\partial {\cal M}, {\tilde g}_{\mu \nu})$ is locally conformally 
mapped to the $D$-dimensional static Einstein cylinder. 
On the conformal boundary $\partial {\cal M}$, we consider a $D$-dimensional static, asymptotically AdS black hole, 
whose spatial section is conformally embedded in the half of the $(D-1)$-sphere--i.e., a spatial section 
of the static Einstein cylinder--as shown by the dashed blue curve in Fig.~\ref{fig1}. 
On such a $D$-dimensional AdS black hole realized in our holographic setting, 
we consider the entanglement entropy for the subregion 
{\rm A} on a spacelike hypersurface $S$ separated by the horizon~(see Fig.~\ref{fig2}). 
The entropy is obtained by tracing over the degrees of freedom located in the subregion {\rm B}.  

From the bulk point of view, there is a static black droplet solution in which the horizon is connected to the boundary horizon of the boundary AdS black hole. We assume that there are no bulk black holes disconnected from the AdS boundaries or black funnels connecting 
two or more boundary horizons.  
According to the Ryu-Takayanagi formula~\cite{Ryu-Takayanagi2006}, the entanglement entropy between A and B is obtained by the 
area of the bulk minimal surface anchored to the bifurcate surface ${\rm C}$ between ${\rm A}$ and ${\rm B}$. 
Since the bulk spacetime is static, there is a bifurcate Killing horizon
in the bulk anchored to the bifurcate surface ${\rm C}$ on the AdS boundary. 
So, the entanglement entropy corresponds to the area of the bifurcate surface of the Killing horizon of the black droplet solution. 

We examine the first law of entanglement entropy in the boundary field theory by applying the Noether charge formula 
to the bulk Einstein gravity in a $(D+1)$-dimensional AlAdS spacetime~\cite{IyerWald1994}. 
Let $\xi^M$ be a Killing vector field with respect to the bulk metric $g_{MN}$.   
The Noether charge $(D-1)$-form ${\bm Q}$ associated with $\xi^M$ is given by 
\begin{align}
\label{Def:Noether_charge}
& {\bm Q}=-\frac{1}{16\pi}\epsilon_{M_1M_2\cdots M_{D-1}NL} \nabla^N \xi^L,  
\end{align}   
where here and hereafter we set the $(D+1)$-dimensional gravitational constant equal to one 
and we omit the indices $M_1 M_2 \cdots $ when writing differential forms, as in the l.~h.~s. above.    
For an arbitrary small deviation ${\delta g}_{MN}$ from the bulk geometry $({\cal M}, g_{MN})$,  
the symplectic potential $D$-form ${\bm \Theta}$ is defined as
\begin{align}
\label{Def:sympletic_form}
& {\bm \Theta}(g,\, \delta g)=\frac{1}{16\pi}\epsilon_{NM_1M_2\cdots M_{D}}g^{NJ}g^{LK}(\nabla_L \delta g_{JK}-\nabla_J \delta g_{LK}). 
\end{align}

As shown in Ref.~\cite{IyerWald1994}, the exterior derivative of the variation of the Noether charge is expressed by the symplectic 
potential ${\bm \Theta}$ as    
\begin{align}
\label{deviation_relation}
d\delta {\bm Q}=d(\xi\cdot {\bm \Theta}), 
\end{align}
where the ``centered dot'' in the r.~h.~s. of Eq.~(\ref{deviation_relation}) denotes the contraction of the 
Killing vector field $\xi^M$ into the first index of the form ${\bm \Theta}$.  
Integrating this over a $D$-dimensional spacelike hypersurface $\Sigma$, we obtain 
\begin{align}
\label{first_law_form}
\int_{\p \Sigma} (\delta {\bm Q}-\xi\cdot {\bm \Theta}(g,\, \delta g))=0, 
\end{align} 
where $\p \Sigma$ is the $(D-1)$-dimensional boundary of the hypersurface $\Sigma$. Applying this equation to a black hole spacetime with 
a bifurcate Killing horizon with respect to $\xi^M$, one can obtain the first law.  
  
In the Fefferman-Graham expansion, the bulk metric is represented by 
\begin{align}
\label{Fefferman_Graham}
& ds^2=g_{MN}dx^M dx^N= \frac{L^2}{z^2} \left(dz^2+ {\tilde g}_{\mu\nu}(z, x)dx^\mu dx^\nu \right),  
\nonumber \\
& {\tilde g}_{\mu\nu}(z, x)={\tilde g}_{(0)\mu\nu}(x)+z^2{\tilde g}_{(2)\mu\nu}(x)+\cdots +z^D{\tilde g}_{(D)\mu\nu}(x)
+\cdots, \quad  
\nonumber \\
& 
x^M =(z, x^\mu) \, \quad (\mu=0, 1, \cdots, D-1),  
\end{align}   
where $L$ is the AdS curvature length.  

For $D=3$ and $D=5$ cases we are interested in, there is no conformal anomaly, and hence there is no 
logarithmic term in (\ref{Fefferman_Graham})~\cite{HaroSkenderis}. Since the background bulk solution $g_{MN}$ is 
static, there is a timelike Killing vector field, $(\p_t)^M$ on ${\cal M}$, and let us take $\xi^M = (\partial_t)^M$. 
We assume that ${\delta g}_{MN}$ is a solution to the linearized equations of motion satisfying 
the Dirichlet boundary condition at the conformal boundary of the bulk AdS ${\cal M}$,  
\begin{align}
\label{Dirichlet}
\delta {\tilde g}_{(0)\mu\nu}=0. 
\end{align} 
Since ${\tilde g}_{(n)\mu\nu}~(0<n<D)$ is the function of ${{\tilde g}}_{(0)\mu\nu}$~\cite{HaroSkenderis}, the leading order of the metric perturbation is 
\begin{align}
\label{leading_(D)coefficient}
{\delta {g}}_{\mu\nu}=O(z^{D-2}). 
\end{align}

When $\p \Sigma$ is the bifurcation surface of a bulk black hole, the second term in Eq.~(\ref{first_law_form}) disappears on $\p \Sigma$ 
and the first term yields the variation of the black hole area~(entropy). In the black droplet solutions, the horizon extends to the AdS 
boundary, and hence the area diverges towards the boundary. Note however that the variation should be finite for the Dirichlet boundary 
condition~(\ref{Dirichlet}). When $\p \Sigma$ is taken as a part of the boundary of $\Sigma$ at the AdS conformal boundary, 
the formula~(\ref{first_law_form}) provides the variation of the holographic energy $E$~\cite{PapadimitriouSkenderis}, defined by 
\begin{align}
\label{holographic_energy}
E=\int_{\p \Sigma} T_{\mu\nu}\xi^\mu n^\nu \,, 
\end{align}
where $T_{\mu\nu}$ is the stress-energy tensor of the boundary field theory, $n^\mu$ is the unit normal vector to $\p \Sigma$, and the volume element of 
$\partial \Sigma$ with respect to the conformal metric $g_{(0)}$ is understood in the integral.  
Applying the holographic stress-energy formula~\cite{HaroSkenderis}, 
$T_{\mu\nu}$ is given in terms of the coefficient ${\tilde g}_{(D)\mu\nu}$ as 
\begin{align}
\label{stress-energy}
T_{\mu\nu}=\frac{DL^{D-1}}{16\pi}{\tilde g}_{(D)\mu\nu} \,, 
\end{align}
in odd dimensional cases.


Our setup for holographic derivation of the first law for entanglement entropy in CFT is as follows. 
As our spatial hypersurface $\Sigma$ for evaluating the Noether charge formula, we take the lower half of the time-slice enclosed by part of the AdS boundary, black droplet horizon and $\chi= \pi/2$ (shaded region in Fig.~\ref{fig1}). 
Our hypersurface $\Sigma$ has the asymptotic boundary $S \subset \p \Sigma$, which itself admits 
its boundary $(D-2)$-sphere denoted by ${\rm D}$ in Fig.~\ref{fig2}. Thus, the hypersurface $\Sigma$ admits the boundaries 
and corners. 
We will impose the reflecting (Neumann) boundary condition at $\chi= \pi/2$ (red dotted line in Fig.~\ref{fig1}). 
We will discuss in Sec.~\ref{sec:4} that this condition is necessary for the purpose of avoiding any additional contributions to the first law from the upper side halfmoon-shaped region. 
In the next section, in more concrete setting, we consider the Noether charge at the boundaries $\partial \Sigma$ 
and also study an additional contribution to the Noether charge from the corner. 
 
\section{The first law of entanglement in the exact droplet solution in $(3+1)$-dimension}
\label{sec:3}
\subsection{The exact droplet solution in $(3+1)$-dimension}
Let us first quickly review the exact solution of a black droplet~\cite{HubenyMarolfRangamani} constructed by the analytic continuation 
of the $(3+1)$-Schwarzschild-AdS solution, 
\begin{align} 
\label{Sch_AdS_4}
& ds_4^2=-F(\rho)d\tau^2+\frac{d\rho^2}{F(\rho)}+\rho^2(d\theta^2+\sin^2\theta d\Phi^2), \nonumber \\
& F(\rho)=\frac{\rho^2}{L^2}+1-\frac{\mu}{\rho}, \quad \mu=\frac{\rho_0^3}{L^2}+\rho_0,  
\end{align}
where $\rho_0$ is the horizon radius. 
The double Wick rotation
\begin{align}
\label{Wick_rotation} 
\tau=i\chi, \qquad \Phi=i\tilde{t}
\end{align}
and the coordinate transformation $\tilde{r}:=\cos\theta$ yields the following warped product type metric including $2$-dimensional de Sitter 
spacetime, 
\begin{align}
\label{two_dS} 
ds^2=F(\rho)d\chi^2+\frac{d\rho^2}{F(\rho)}+\rho^2\left(-(1-\tilde{r}^2) d\tilde{t}^2+\frac{d\tilde{r}^2}{1-\tilde{r}^2}\right).   
\end{align}
By the rescaling of the coordinates $(\chi, \tilde{r}, \tilde{t})$, 
\begin{align}
\label{coordinate_rescaling} 
\tilde{t}=\frac{r_0}{l^2}t, \qquad \tilde{r}=\frac{r_0}{r}, \qquad \chi=\frac{L}{l}r_0\varphi, 
\end{align} 
one obtains the following bubble solution 
\begin{align} 
\label{BTZ_BD}
& ds^2=\frac{d\rho^2}{F(\rho)} 
+\frac{\rho^2r_0^2}{l^2r^2}\left[-\frac{r^2-r_0^2}{l^2}dt^2+\frac{l^2dr^2}{r^2-r_0^2}
+\frac{L^2r^2F(\rho)}{\rho^2}d\varphi^2  \right], 
\end{align}
where $\varphi$ circle has a period $2\pi$. The $S^1$ circle along $\p_\varphi$ shrinks to zero size 
at the bubble radius, $\rho=\rho_0$, i.~e.~, $F(\rho_0)=0$. By imposing a regularity condition at the bubble radius, 
$r_0$ is expressed by $\rho_0$ as  
\begin{align} 
\label{r0_rho0_relation}
r_0=\frac{2\rho_0lL}{3\rho_0^2+L^2}. 
\end{align}
Note that the boundary metric (i.e., $3$-dimensional metric in the square brackets in Eq.~(\ref{BTZ_BD})) is conformal to the static BTZ black hole with the horizon radius $r=r_0$~\cite{BTZblackhole}. 
Note also that in the $4$-dimensional bulk, there is a bulk horizon which extends from the bubble radius $\rho_0$, deep inside the bulk spacetime to the boundary BTZ black hole horizon. 
In this way, this bulk metric describes a black droplet as illustrated in Fig.~\ref{fig1}. 

Let us change the coordinates in Eq.~(\ref{BTZ_BD}) into 
the Fefferman-Graham coordinate system~(\ref{Fefferman_Graham}) by 
\begin{align} 
\label{z-rho_BTZ}
& \frac{1}{\rho(\eta, \, z)}=z\left(\frac{1}{\eta L}+\alpha_2(\eta)z^2+\alpha_3(\eta)z^3+\cdots    \right), \nonumber \\
& r(\eta, \, z)=r_0\left(\eta+\beta_2(\eta)z^2+\beta_4(\eta)z^4+\cdots \right), \nonumber \\
& \alpha_2=-\frac{\eta^2-2}{4L\eta^3}, \qquad \alpha_3=-\frac{\mu}{6L^2\eta^4}, \nonumber \\
& \beta_2=\frac{\eta^2-1}{2\eta}, \qquad \beta_4=\frac{\eta^4-1}{8\eta^3}, \cdots. 
\end{align}
Then, we obtain the stress-energy tensor~(\ref{stress-energy}) on the boundary field theory by 
\begin{align}
\label{stress-energy_BTZ} 
& T_{tt}=-\frac{\mu Lr_0^2(\eta^2-1)}{16\pi  l^4\eta^3}, \nonumber \\
& T_{\eta\eta}=\frac{\mu L}{16\pi \eta^3(\eta^2-1)}, \nonumber \\
& T_{\varphi\varphi}=-\frac{\mu Lr_0^2}{8\pi  l^2\eta}.  
\end{align}
Since the BTZ black hole is in odd-dimensions, the trace anomaly vanishes, i.e.,~${T_\mu^\mu}=0$, as seen 
in Eq.~(\ref{stress-energy_BTZ}). So, while even-dimensional black droplet solutions have infinite energy due to the 
anomaly~\cite{MarolfSantos2019}, the holographic energy $E$ in Eq.~(\ref{holographic_energy}) converges to a finite value. 
When one applies the formula~(\ref{first_law_form}) to the black droplet solution~(\ref{BTZ_BD}), one should note that the $2$-dimensional surface 
$\p \Sigma$ consists of three parts: a part at the AdS boundary, $z=0$, a part of the bulk bifurcate Killing horizon, $r=r_0$, and a part of the timelike surface 
at $r=\infty$. In the next subsection we evaluate Eq.~(\ref{first_law_form}) on part of the timelike surface at infinity.

\subsection{Perturbation of the $(3+1)$-dimensional black droplet solution}
\label{subsec:pert3bdrplt}

We consider metric perturbations $h_{MN} := \delta g_{MN}$ of the black droplet solution~(\ref{BTZ_BD}). 
For convenience we express our coordinate system as $x^M = (y^a, x^i)$ with $y^a:= (\rho, \varphi), \: x^i:=(t,r)$ and denote by  
$\tilde{\gamma}_{ij}$ the $2$-dimensional de Sitter metric spanned by $x^i$ in (\ref{BTZ_BD}).  
We can impose the following gauge condition:
\begin{align}
 h_{ai} =0 \,, \quad h_{ij} \propto {\tilde \gamma}_{ij} \,. 
\label{gauge_choice_two}
\end{align}
This condition may immediately be justified by, e.g., applying the scalar-type perturbations of~\cite{KodamaIshibashi2003} to the $4$-dimensional 
Schwarzschild-AdS solution~(\ref{Sch_AdS_4}) with the scalar harmonics $\mS_{\bf k}(x^i)$ and taking the gauge $f_a=0,\: H_T =0$ in the terminology of~\cite{KodamaIshibashi2003}. Note however that 
when applying the formalism of~\cite{KodamaIshibashi2003} to the present case, the scalar ``harmonics" $\mS_{\bf k}(x^i)$ corresponds to 
the solutions of the Klein-Gordon equation on the Lorentzian sphere (i..e., de Sitter space with the metric ${\tilde \gamma}_{ij}$), 
rather than the standard eigen-function of the laplace operator with respect to the unit two-sphere metric. 

Now we assume that our perturbations $h_{MN}$ satisfy the symmetry along $\p_\varphi$. Under the gauge choice~(\ref{gauge_choice_two}), 
the variation of the Noether charge~(\ref{Def:Noether_charge}) and the symplectic two 
form~(\ref{Def:sympletic_form}) for the Killing vector $\xi=\p_t$ are calculated 
for $t=\mbox{const.}$ and $r=\infty$ surface for the metric~(\ref{BTZ_BD}) as 
\begin{align}
\label{Noether_charge_two} 
& \int_{r=\infty}\delta {\bm Q}=\frac{1}{32\pi}\int \epsilon_{\rho\varphi tr}(h_a{}^a - h_i{}^i ) g^{tt}g^{rr} \partial_rg_{tt} 
-\frac{1}{16\pi}\int \epsilon_{\rho\varphi t r}g^{tt}g^{rr}\left( \partial_t{h}_{rt}- \partial_r{h}_{tt}\right), 
%
\end{align}
and 
\begin{align} 
\label{sympletic_two} 
& \int_{r=\infty} \xi\cdot \Theta=
-\frac{1}{16\pi}\int \epsilon_{\rho\varphi tr} g^{rr}g^{tt}
\Biggl[\frac{1}{2} h_i{}^i  \partial_r g_{tt}
-g_{tt} \partial_r h_a{}^a + \partial_t{h}_{rt}- \partial_r {h}_{tt}\Biggr],  
\end{align}
where we denote the partial traces by $h_a{}^a = h_\rho{}^\rho + h_\varphi{}^\varphi,\: h_i{}^i =h_t{}^t+h_r{}^r$. 
Note that as usual, the indices of $h_{\mu \nu}$ are raised and lowered by the background metric $g_{\mu \nu}$ and its inverse, e.g., 
$h_\mu{}^\nu := h_{\mu \lambda} g^{\lambda \nu}$. Then, the combination of Eq.~(\ref{Noether_charge_two}) and (\ref{sympletic_two}) yields 
\begin{align}
\label{Combination_two}
& \int_{r=\infty}(\delta {\bm Q}-\xi\cdot \Theta)=\frac{1}{16\pi}\int \epsilon_{\rho\varphi t r}
g^{rr}g^{tt}\Biggl[\frac{1}{2} h_a{}^a \partial_rg_{tt} 
-g_{tt} \partial_r h_a{}^a \Biggr]. 
\end{align}
With the condition (\ref{gauge_choice_two}) and the perturbed vacuum Einstein equations, we can find that the partial trace $h_a{}^a$ 
vanishes in $4$-dimensional bulk~\cite{KodamaIshibashi2003}, i.e.,  
\begin{align}
\label{trace_zero}
h_a{}^a 
= {h_\varphi}^{\varphi}+ {h_\rho}^{\rho}=0. 
\end{align} 
Thus, the r.h.s. of Eq.~(\ref{Combination_two}) vanishes  
on $t=\mbox{const.}$ and $r=\infty$ surface. It follows from the holographic interpretation \cite{CHM2011} that 
the first law of the entanglement entropy $S$ is satisfied:  
\begin{align}
\label{first_law}
\delta E=T\delta S
\end{align}
in the black droplet solution, where $T$ denotes the temperature computed from the surface gravity of the bifurcate Killing horizon.  
This result is peculiar to the $4$-dimensional bulk case, because Eq.~(\ref{trace_zero}) is 
satisfied only for $D=3$ case. In the other dimensions, we need to impose boundary conditions for bulk metric perturbations 
to derive the first law, as we will see in the next section. 
%

\section{The first law in $(4+1)$-dimensional field theory in AdS black hole background}
\label{sec:4}

In this section, we examine the first law of the entanglement entropy for CFT in $D=5$-dimensions by using the holographic principle and applying the Noether charge formula as in the previous section. For the use of the holographic ideas, it would be most convenient if there is, as the gravity dual, any known exact solution of the $(5+1)$-dimensional black droplet that possesses the boundary $D=5$-dimensional metric 
\begin{align} 
\label{D=5_Sch_AdS} 
ds_5^2 =- f(r)dt^2+\frac{dr^2}{f(r)}+r^2 d\Omega_3^2 \,, \quad f(r):= 1 - \frac{M}{r^2}  + \frac{r^2}{L^2} \,, 
\end{align}
where $d\Omega_3^2:= \sigma_{m n} d\theta^m d\theta^n$ is the metric of $3$-dimensional unit sphere with angular coordinates $\theta^m~(m=1,2,3)$. 
Unfortunately there is no such an analytic solution available in $(5+1)$-dimension. But, fortunately, with the Noether charge formula at our hand, all we need for our purpose is the asymptotic behavior of the gravity dual near the boundary rather than the entire structure of the explicitly written exact solution. In fact, in the $D=3$ case examined in~\ref{subsec:pert3bdrplt}, we have the exact gravity dual (\ref{BTZ_BD}) available, but 
we need not use the explicit expression of (\ref{BTZ_BD}) to derive the first law~(\ref{first_law}) for the boundary CFT.  
Therefore, in this section, we simply assume that the desired gravity dual exists. We will attempt to justify this assumption 
by perturbatively constructing a $6$-dimensional black droplet solution in the next section.


Since our $5$-dimensional boundary (\ref{D=5_Sch_AdS}) itself is an asymptotically AdS spacetime with a $4$-dimensional timelike 
surface as its conformal boundary (that is, $r \rightarrow \infty$ in (\ref{D=5_Sch_AdS}): see point D in Fig.~\ref{fig2} and also compare $\rho=\infty, \chi= \pi/2$ in Fig.~\ref{fig1}), 
by analyzing the asymptotic behavior of the gravity dual, we can holographically study the asymptotic fall-off behavior of our $5$-dimensional 
CFT at large distances $r \rightarrow \infty$. 
By doing so we can estimate the energy flux---if it exists---across the $4$-dimensional boundary $r \rightarrow \infty$ and discuss conditions---if necessary---for the first law to hold. 
In Subsec.~\ref{subsec:energyflux}, we holographically study possible energy flux at the boundary $r \rightarrow \infty$ and derive 
the condition for having no energy flux across the boundary. For this purpose, it is sufficient to examine the metric perturbations of the $(5+1)$-dimensional pure AdS spacetime since the asymptotic behavior of the gravity dual near the boundary ($\rho \to \infty, \chi \to \pi/2$ of Fig.~\ref{fig1}) should be approximately well described by metric perturbations of the pure AdS background spacetime near the corresponding boundary (see also Eq.~(\ref{global_AdS}) below).  Then, 
in Subsec.~\ref{subsec:firstlaw}, by considering metric variations of the $(5+1)$-dimensional static black droplet---which is   
assumed to exist as the gravity dual, just mentioned above---we derive the first law under the no flux condition.     
%

\subsection{The energy flux on the perturbed AdS geometry}
\label{subsec:energyflux}
We consider, as our background, the $(5+1)$-dimensional AdS bulk with the metric of the form:  
\begin{align} 
\label{global_AdS}
& ds_6^2 = - F_0(\rho) dt^2 + \frac{d \rho^2}{F_0(\rho)} +
 \rho^2 d\Omega_{4}^2
 \,, \quad F_0(\rho) := 1 + \frac{\rho^2}{L^2} \,,  
\end{align}
where $d \Omega_{4}^2$ denotes the standard metric of the $4$-dimensional unit sphere. 
We use the perturbation formalism of~\cite{IshibashiWald2004}, which is most convenient for the current purpose and is also useful in the next section. 
Accordingly we use coordinates $y^a := (t,\rho)$ to  express the $2$-dimensional AdS metric $g_{ab}$ given by the first two terms in (\ref{global_AdS}), 
and angular coordinates $x^i =( \chi, \theta^m) $ with $0 \leq \chi \leq \pi$ to the metric $\gamma_{ij}$ of the $4$-dimensional unit sphere $\Omega_{(4)}$:  
\begin{equation}
 d \Omega_{4}^2 = \gamma_{ij} dx^i dx^j := d\chi^2 + \sin^2 \! \chi d\Omega_{3}^2 = d\chi^2 + \sin^2\! \chi \sigma_{mn} d\theta^m d\theta^n  \,. 
\end{equation}
Note that the coordinate $\chi$ is later changed to $r$.

On this background, we consider the metric perturbations, which can in general be classified into three types: the scalar-, the vector-, and the tensor-type according to their tensorial behavior on the $4$-dimensional sphere $\Omega_{(4)}$ as explained in~\cite{IshibashiWald2004}. 
The tensor-type perturbations are characterized as the transverse and trace-free part of $\delta g_{ij}$, and in particular, $\delta g_{t \chi}=0$,  
and therefore there is no contribution to the energy flux $T_{tr}$ from the tensor-type perturbations. We discuss the vector-type perturbations 
in Appendix. In the following we focus on the scalar-type metric perturbations. 

It is convenient to introduce the harmonic functions $\mS_{(l)}$ 
on the unit four-sphere $\Omega_{(4)}$, which satisfy  
\begin{align} 
\label{4dim_harmonic}
D^iD_i {\mS}_{(l)}+l(l+3)\mS_{(l)}=0, \quad l=0, 1, 2, \cdots,  
\end{align}
with $D_i$ being the covariant derivative with respect to the metric $\gamma_{ij}$.  
Then the scalar-type metric perturbations can be expressed in terms of $\mS_{(l)}$ as
\begin{align} 
\label{6D_pert_variable}
& \delta g_{ab}= \sum_{ l} H_{(l)ab}\mS_{(l)}, \qquad \delta g_{ai}= \sum_{ l}  H_{(l)a}D_i\mS_{(l)} , 
\nonumber \\
& \delta g_{ij}= \sum_{ l} H_{(l)L}\gamma_{ij}\mS_{(l)}+H_{(l)T}\left(D_iD_j-\frac{1}{4}\gamma_{ij}D^kD_k   \right) \mS_{(l)},  
\end{align}
where $H_{(l)ab}$, \, $H_{(l)a}$, $H_{(l)L}$, and $H_{(l)T}$ are 
functions of the coordinates $y^a=(t, \rho)$.  
Hereafter, for brevity we omit the indices $(l)$ labeling the mode, unless otherwise stated. 
We introduce the following perturbation variables $Z$ and $Z_{ab}$:  
\begin{align}
\label{gauge_invariant_Z}
& Z=4\left(H_{L}+\frac{l(l+3)}{4}H_{T}+2\rho(D^a\rho)X_a\right), \nonumber \\
& Z_{ab}=\rho^2(H_{ab}+D_aX_{b}+D_bX_{a})+\frac{3}{4}Zg_{ab}, \nonumber \\
& X_{a}:=-H_{a}+\frac{1}{2}\rho^2D_a\left(\frac{H_{T}}{\rho^2} \right),  
\end{align}
where $D_a$ is the covariant derivative with respect to the $2$-dimensional unperturbed metric $g_{ab}$. 
Note that the two variables $Z$ and $Z_{ab}$ are gauge-invariant. 
The perturbed Einstein equations can be expressed in terms of $Z$ and $Z_{ab}$ and furthermore 
reduced to the following equation for a single scalar field $\Phi$ on the $2$-dimensional AdS spacetime with the metric $g_{ab}$:  
\begin{align} 
\label{master_eq}
D^aD_a\Phi-\frac{2+l(l+3)}{\rho^2}\Phi=0, 
\end{align} 
in terms of which $Z$ and $Z_{ab}$ are expressed as
\begin{align}
\label{Z_master_relation} 
& Z_{ab}=\left(D_aD_b-\frac{1}{L^2}g_{ab}\right)(\rho^2\Phi), \nonumber \\
& Z={Z_a}^a=\left(D^aD_a-\frac{2}{L^2}\right)(\rho^2\Phi). 
\end{align}

Now, using the gauge degrees of freedom, we can choose the following gauge
\begin{align} 
\label{6dim_gauge_choice_dynamical}
H_{\rho}=H_{\rho\rho}=H_{t\rho}=0 \,. 
\end{align}
In this gauge, $X_{\rho}$, $X_{t}$, and $H_{T}$ are determined 
by the second and third lines in Eqs.~(\ref{gauge_invariant_Z}) as 
\begin{align} 
\label{XHT_6dim_dynamical}
& X_{\rho}=\frac{1}{2\sqrt{F_0}} \int \frac{d\rho}{\rho^2} \left( {\sqrt{F_0 }} Z_{\rho\rho}-\frac{3Z}{4\sqrt{F_0} } \right) , 
\nonumber \\
& X_{t}= F_0 \int \frac{d\rho}{\rho^2 F_0} \left( {Z_{t \rho}}+i\omega \rho^2 X_{\rho} \right) , \nonumber \\
& H_{T}=2\rho^2\int \frac{d\rho}{\rho^2}  {X_{\rho}} ,  
\end{align}
where as previously introduced $F_0 = 1 + \rho^2/L^2$. 
From the third line of Eqs.~(\ref{gauge_invariant_Z}), $H_{t}$ is expressed as 
\begin{align} 
\label{Ht_6dim_dynamical}
H_{t}=-X_{t}-\frac{1}{2}i\omega H_{T},  
\end{align}
where we have assumed that all the metric functions depend on $t$ as $e^{-i\omega t}$.  

Let us expand the master variable $\Phi$ near the AdS boundary $\rho \rightarrow \infty$ as 
\begin{align} 
\label{CapitalPhi}
\Phi =a_{0}+\frac{a_{1}}{\rho}+\frac{a_{2}}{\rho^2} + \cdots.  
\end{align}
The boundary condition~(\ref{Dirichlet}) imposes $a_{0}=0$ and sets the integral constants in Eqs.~(\ref{XHT_6dim_dynamical}) to 
zero. Then the stress-energy tensor on the boundary theory~(\ref{stress-energy}) is proportional to the coefficient $a_{1}$. 

The perturbed metric is transformed to the Fefferman-Graham coordinate~(\ref{Fefferman_Graham}) by 
\begin{align} 
\label{global_AdS_FG}
& \rho=\frac{\sqrt{1+ {r^2}/{L^2}}}{z} 
\Biggl[1+\frac{L^2(L^2-r^2)}{4(L^2+r^2)} z^2
+\frac{L^4(L^4-4L^2r^2+r^4)}{16(L^2+r^2)^2}z^4+\cdots \Biggr], \nonumber \\
& L\tan\chi=r\frac{4+L^2z^2}{4-L^2z^2}.  
\end{align}   
Then, expanding the metric component ${\delta g}_{tr}$ as a series in $z$, 
$T_{tr}$ in~(\ref{stress-energy}) can be read from the coefficient of $z^3$ as
\begin{align}
\label{t_eta_component} 
T_{tr}\sim \frac{a_{1} }{ \left(1+ r^2/L^2 \right)^{3/2} } \partial_r\mS. 
\end{align} 
Therefore in the boundary theory side, the energy flux across the AdS boundary $r=\infty$ (see $\chi \rightarrow \pi/2$ in Fig.~\ref{fig1}) becomes 
\begin{align}
\label{energy_flux} 
J\sim \lim_{r\to \infty}\int T_{t\mu}n^\mu\sqrt{1+\frac{r^2}{L^2}}r^3d\Omega_3\,dt
\sim a_{1}\lim_{r\to \infty} r^2 \partial_r \mS,  
\end{align} 
where $n^\mu$ is the unit normal to the timelike AdS boundary hypersurface at $r=\infty$. 
From the harmonic equation~(\ref{4dim_harmonic}), 
$\mS$ behaves near the boundary as 
\begin{align} 
 \mS  \simeq A+\frac{B}{r}.  
\end{align}
By Eq.~(\ref{energy_flux}), the condition for no energy flux is 
\begin{align}
\label{no_energy_flux}
B=0. 
\end{align} 
In terms of the angle coordinate $\chi$, this condition is equivalent to imposing  
\begin{align}
\label{boundary_con_S_Der}
\lim_{\chi\to {\pi}/{2}} \partial_\chi \mS =0 \,, 
\end{align} 
for the harmonic functions. It is easily checked from Eq.~(\ref{4dim_harmonic}) that this is satisfied only for even $l$. 
In the next subsection, we will check that the first law of the entanglement entropy is satisfied under the above no-energy flux conditions.

\subsection{The first law of the $6$-dimensional black droplet solution}
\label{subsec:firstlaw}
As mentioned before, throughout this section, we assume that there is a static black droplet solution whose conformal boundary 
is the $D=5$ asymptotically AdS, static $SO(4)$ symmetric black hole with the metric $g_{\mu \nu}(z=0, x)$ given 
by (\ref{D=5_Sch_AdS}). 
Such a black droplet solution should admit the Fefferman-Graham expansion~(\ref{Fefferman_Graham}) with $D=5$ metric 
$g_{\mu \nu}(z,x)$ conformal to the boundary metric (\ref{D=5_Sch_AdS}).  
In this subsection, as the radial coordinate in the $D=5$-boundary, we use $\chi$ instead of $r$, so that $x^\mu =(t, \chi, \theta^m)$.

Now, to derive the first law applying the Noether charge formula, let us consider metric perturbations $h_{\mu \nu} = \delta g_{\mu \nu}$. 
For simplicity, we assume that the perturbed metric satisfies $SO(4)$ symmetry, in which the only off-diagonal 
term appeared is $h_{t\chi}$. 
Since our background boundary metric is the $5$-dimensional static black hole, such a $SO(4)$ symmetric perturbation corresponds to monopole or mass perturbation for the boundary black hole. We briefly comment on possible contributions to the Noether charge formula 
from general perturbations without $SO(4)$ symmetry in Appendix. 

On our assumptions, the Noether charge~(\ref{Def:Noether_charge}) and the symplectic form~(\ref{Def:sympletic_form}) 
are evaluated at $t=\mbox{const}.$ and $\chi=\pi/2$ hypersurface as 
\begin{align}
\label{Noether_charge_D5}
& \int_{\chi=\pi/2}\delta {\bm Q}=-\frac{1}{16\pi}\int \epsilon_{z123 t\chi}g^{tt}g^{\chi\chi}
 \Biggl[
          \partial_t h_{\chi t} - \partial_\chi h_{tt}  
 +(\partial_\chi g_{tt}) 
\Bigl( h_t{}^t + h_\chi{}^\chi - \frac{1}{2} h_\mu{}^\mu\Bigr) \Biggr],  
\end{align}
\begin{eqnarray}
\label{Sympletic_D5}
\int_{\chi=\pi/2}[\xi\cdot \Theta]_{z123}
&=&-\frac{1}{16\pi}\int \epsilon_{z123 t\chi}g^{\chi\chi}
\Biggl[g^{tt} \partial_t h_{\chi t} + \frac{1}{2} h_t{}^t g^{tt} \partial_\chi g_{tt} 
  -\frac{1}{2}h_m{}^lg^{mn} \partial_\chi g_{nl}   
\nonumber \\ 
&{}& \, 
  + \frac{1}{2} h_\chi{}^\chi (g^{tt} \partial_\chi g_{tt} + g^{mn} \partial_\chi g_{mn})
  - g^{tt} \partial_\chi h_{tt} 
-g^{mn} \partial_\chi h_{mn} + h_m{}^n g^{ml} \partial_\chi g_{nl} \Biggr],  
\end{eqnarray}
where we emphasize that the indices $l,m,n$ represent the angular coordinates on the unit three-sphere: $\theta^m \in \Omega_3$.  
Each term includes the off-diagonal term $\delta g_{t\chi}$, but this vanishes in the combination of 
Eq.~(\ref{Noether_charge_D5}) and Eq.~(\ref{Sympletic_D5}) as follows, 
\begin{eqnarray}
\label{comb_Q_Sym_D5}
 \int_{\chi=\pi/2}(\delta {\bm Q}-\xi\cdot \Theta)
  &=&\frac{1}{16\pi}\int \epsilon_{z123 t\chi} g^{\chi \chi} \Biggl[
-\frac{1}{2}g^{tt} h_m{}^m \partial_\chi g_{tt} 
 -\frac{1}{2} g^{mn} h_m{}^l \partial_\chi g_{nl} 
-\frac{1}{2} h_\chi{}^\chi g^{mn} \partial_\chi g_{mn} + g^{mn} \partial_\chi h_{mn} 
\Biggr]. 
\end{eqnarray}
From the expression above~(\ref{comb_Q_Sym_D5}), we immediately find that 
if both the background and perturbed metrics satisfy the Neumann boundary condition 
\begin{align}
\label{Neumann_bc_5D}
\lim_{\chi\to {\pi}/{2}} \partial_\chi g_{\mu\nu}=\lim_{\chi\to {\pi}/{2}} \partial_\chi h_{\mu\nu} =0, 
\end{align} 
then Eq.~(\ref{comb_Q_Sym_D5}) vanishes. This also shows the first law of the entanglement entropy~(\ref{first_law}) in the $D=5$ case, 
as in the $D=3$ case. 

As shown in the previous subsection, the latter condition corresponds to 
the no-energy flux condition~(\ref{boundary_con_S_Der}) or (\ref{no_energy_flux}) for the scalar-type perturbations.  
As for the vector- and tensor-type perturbations, it is not obvious from the above expression alone that one has no contributions to (\ref{comb_Q_Sym_D5}). 
Nevertheless, one can show, at least for the present $SO(4)$ symmetric case, 
that the energy flux caused by the perturbation $h_{\mu \nu}(z, x)$ vanishes on $\chi=\pi/2$ as follows: 
The perturbed boundary stress energy tensor $\delta T_{\mu\nu}$ satisfies 
the conservation law~\cite{HaroSkenderis}
\begin{align}
\label{conservation_law}
\nabla_\nu {\delta T_\mu}^\nu=0, 
\end{align}
where $\nabla_\mu$ is the covariant derivative with respect to the conformal boundary metric ${\tilde g}_{(0)\mu\nu}$ in 
Eq.~(\ref{Fefferman_Graham}) under the boundary condition~(\ref{Dirichlet}). 
The $\chi$-component of (\ref{conservation_law}) is written as  
\begin{align}
\label{conservation_law_chi}
\p_\nu \delta {T_\chi}^\nu-{\Gamma^\alpha}_{\nu\chi}\delta {T_\alpha}^\nu
+{\Gamma^\nu}_{\nu\alpha}\delta {T_\chi}^\alpha=0.  
\end{align} 
Since the only nonzero off-diagonal term of $\delta {T_\mu}^\nu$ is  $\delta {T_\chi}^t$~(or $\delta {T_t}^\chi$), and 
the background metric is static diagonal as assumed, the last two terms in Eq.~(\ref{conservation_law_chi}) must 
vanish on $\chi=\pi/2$ by the boundary condition~(\ref{Neumann_bc_5D}). Thus, Eq.~(\ref{conservation_law_chi}) leads 
to  
\begin{align}
\label{conservation_law_chi_II}
\lim_{\chi\to \frac{\pi}{2}}\nabla_\nu {\delta T_\chi}^\nu=
\lim_{\chi\to \frac{\pi}{2}}(\p_t \delta {T_\chi}^t+\p_\chi \delta {T_\chi}^\chi)=0.  
\end{align} 
The boundary condition~(\ref{Neumann_bc_5D}) implies 
$\lim_{\chi\to {\pi}/{2}} \partial_\chi {\tilde g}_{\mu\nu}=\lim_{\chi\to {\pi}/{2}} \partial_\chi {\tilde h}_{\mu\nu} =0$. 
The fifth-order of the Fefferman-Graham expansion of the perturbed metric ${\tilde h}_{(5)\mu \nu}
= \delta {\tilde g}_{(5)\mu \nu}$ satisfies $\partial_\chi {\tilde h}_{(5)\mu\nu}=0$ at $\chi=\pi/2$, and therefore it follows from the holographic stress-energy tensor formula~(\ref{stress-energy}) that $\lim_{\chi\to {\pi}/{2} }\p_\chi \delta {T_\chi}^\chi=0$. Then, since $\p_t {\tilde h}_{\mu \nu}=-i\omega {\tilde h}_{\mu\nu}$ 
on the linearized perturbation, Eq.~(\ref{conservation_law_chi_II}) yields
\begin{align}
\lim_{\chi\to {\pi}/{2}}\delta {T_\chi}^t=0. 
\end{align}
This is nothing but the no energy flux condition, as observed in Eq.~(\ref{no_energy_flux}).

\section{The perturbative approach in $(5+1)$-dimensional black droplet solution}
\label{sec:5}
In this section, we will construct the asymptotic geometry of the $6$-dimensional black droplet solution under the boundary condition~(\ref{Neumann_bc_5D}) 
in order to justify our analysis based on the holographic method in the last section. 
We will also examine the energy density of the boundary field theory. When the boundary black hole mass 
parameter $M$ in Eq.~(\ref{D=5_Sch_AdS}) is small compared with the AdS scale, i.e., $M\ll L^2$, the 
asymptotic metric can be constructed from the static perturbation of the pure AdS spacetime~(\ref{global_AdS}). 
As shown below, a negative energy appears near the asymptotic region, inconsistent with 
the $D=3$ result~(\ref{stress-energy_BTZ}). We also show that the holographic energy~(\ref{holographic_energy}) is 
finite. This is because there is no trace anomaly in odd dimensions, being different from the even 
dimensional case~\cite{MarolfSantos2019}.   

We require that the conformal boundary metric of the perturbed bulk metric coincides with the 
metric~(\ref{D=5_Sch_AdS}) asymptotically in the sense that  
\begin{align} 
\label{asymptotic_D=5_Sch_AdS}
& ds_5^2=- f(r)dt^2
+ \frac{1}{f(r)} {(1+O(r^{-6}))} dr^2  
 + r^2\left(1+O(r^{-6}) \right)d{\Omega_3}^2 \,, 
\end{align}
where $f(r) = 1 - {M}/{r^2}  + {r^2}/{L^2}$ as previously introduced by (\ref{D=5_Sch_AdS}).

In the static perturbation with $\omega =0$, $Z_{t\rho}=H_{t\rho}=H_{t}=X_t=0$. Adopting the gauge choice
\begin{align} 
\label{6dim_gauge_choice}
H_{\rho}=H_{\rho\rho}=0, 
\end{align}
$X_{\rho}$, $H_{T}$ are determined by the gauge-invariant 
variables $Z_{\rho\rho}$, $Z$ by Eqs.~(\ref{XHT_6dim_dynamical}). One can always set  
\begin{align}
\label{H_tt_static_5D}
\lim_{\rho\to \infty}\frac{H_{tt}}{\rho^2}=0
\end{align}
by choosing the integral constant in the first line in Eqs.~(\ref{XHT_6dim_dynamical}) appropriately. 
To construct the boundary metric (\ref{asymptotic_D=5_Sch_AdS}) by perturbing the bulk metric~(\ref{global_AdS}),   
we find that the angle coordinate $\chi$ in Eq.~(\ref{global_AdS}) and the radial coordinate $r$ of the boundary metric (\ref{asymptotic_D=5_Sch_AdS}) should be related as
\begin{align} 
\label{chi_xi_Tr}
d\chi= \frac{dr}{Lf(r)} \,. 
\end{align}
This is integrated and approximately expressed at large $r$ as 
\begin{align} 
\label{chi_xi_relation}
\tan\chi=\frac{r}{L}\left(1-\frac{ML^2}{5r^4}+\frac{3ML^4}{35r^6}+\cdots  \right). 
\end{align}
Then, $h_{rr}=\delta g_{rr}$ at $O(\rho^2)$ can be expanded as 
\begin{align}
\label{expansion_rr}
\lim_{\rho\to \infty}\frac{ h _{rr}}{\rho^2} = \frac{1}{L^2 f(r)^2} \sum_{k=0}\frac{\alpha_k}{r^{2k}}.
\end{align} 
Therefore the leading order of the bulk perturbed metric with $O(\rho^2)$ becomes 
\begin{align}
\label{def_h}
& ds_6^2\simeq\frac{d\rho^2}{F_0(\rho)}+\frac{\rho^2}{L^2 f(r)}\Biggl[
\left(1+\sum_{k=0}\frac{\alpha_k}{r^{2k}} \right) \frac{dr^2}{f(r)} 
+ f(r)\left\{-dt^2+L^2(\sin^2\chi+ h_L) d\Omega_3^2\right\}\Biggr]+O(\rho^0), 
%
\end{align}
where $h_L$ is the function of $\chi$, arising from the perturbation. 
If both the conditions 
\begin{align}
\label{cond_alpha}
\alpha_0=\alpha_1=\alpha_2=0, 
\end{align}
and 
\begin{align}
\label{cond_h}
f(r)L^2 (\sin^2\chi+h_L)= r^2+O(r^{-4}), 
\end{align}
are satisfied, our bulk metric is conformal to the $5$-dimensional Schwarzschild-AdS metric~(\ref{asymptotic_D=5_Sch_AdS}), 
up to $O(r^{-4})$. As shown below, both the conditions~(\ref{cond_alpha}) and (\ref{cond_h}) are satisfied for a suitable 
superposition of the first three even mode harmonic functions, $\mS_{(l)}~(l= 2, 4, 6)$ in Eq.~(\ref{4dim_harmonic}), which satisfy 
the boundary condition~(\ref{Neumann_bc_5D}). From now on we show the indices $(l)$ labeling the mode.  

The analytic solution of Eq.~(\ref{master_eq}) that satisfies the regularity at $\rho=0$ is given by the hypergeometric function $F$ with the 
argument $F_0(\rho)= 1+\rho^2/L^2$ (see \cite{IshibashiWald2004}):  
\begin{eqnarray}
\label{hypergeometric}
 \Phi_{(l)} &=&a_{0(l)} \left[ \frac{ F_0(\rho)-1}{F_0(\rho) }\right]^{1 + {l}/{2}} \cdot 
        \Biggl\{ F\left(\frac{l+2}{2}, \frac{l+2}{2}, \frac{1}{2}, \frac{1}{F_0(\rho)} \right) 
\nonumber \\
&{}& \qquad \qquad \qquad \qquad \qquad
-2 \left[ \frac{\Gamma\left(\frac{l+3}{2}\right)}{\Gamma\left(\frac{l+2}{2}\right)} \right]^2 
         \frac{1}{\sqrt{F_0(\rho)}} \cdot F\left(\frac{l+3}{2}, \frac{l+3}{2}, \frac{3}{2},  \frac{1}{F_0(\rho)} \right)
       \Biggr\}. 
\end{eqnarray}
This equation enables us to describe $a_{1(l)}$ by $a_{0(l)}$ as follows, 
\begin{align}
\label{a1_a0_relation}
a_{1(l)}=-2L \left[ \frac{\Gamma\left(\frac{l+3}{2}\right)}{\Gamma\left(\frac{l+2}{2}\right)} \right]^2 a_{0(l)}. 
\end{align} 
The first three even mode harmonic functions $\mS_{(l)}$ are given by  
\begin{align}
\label{harmonic_four}
& \mS_{(2)}(\chi)=\frac{3}{4}(5\cos^2\chi-1), \nonumber \\
& \mS_{(4)}(\chi)=\frac{5}{8}(1-14\cos^2\chi+21\cos^4\chi), 
\nonumber \\
& \mS_{(6)}(\chi)=\frac{7}{64}(-5+135\cos^2\chi-495\cos^4\chi+429\cos^6\chi). 
\end{align}
By Eqs.~(\ref{XHT_6dim_dynamical}), the leading term of $H_{(l)T}$ becomes 
\begin{align}
\label{integrate_con_c}
H_{(l)T}=c_{(l)}\rho^2+O(1), 
\end{align}
where $c_{(l)}$ is the integral constant for $H_{(l)T}$ in Eqs.~(\ref{XHT_6dim_dynamical}).  
Then, from (\ref{6D_pert_variable}), we find that $\delta g_{\chi\chi}$ is asymptotically, $\rho \rightarrow \infty $, described by  
\begin{align}
\label{g_chichi}
 h_{\chi\chi}
&=\sum_{l=2, 4, 6} \Biggl[\left(H_{(l)L}+\frac{l(l+3)}{4}H_{(l)T}\right)\mS_{(l)}(\chi) 
+H_{(l)T}\partial_\chi^2\mS_{(l)}(\chi)\Biggr]
\nonumber \\
& 
=\sum_{l=2,4,6} \left(\frac{3a_{0(l)}}{L^2}\mS_{(l)}+c_{(l)}\p_\chi^2\mS_{(l)}\right)\rho^2+O(1), 
\end{align} 
where Eq.~(\ref{4dim_harmonic}) is used in the first line. As for the second line, we first write $Z_{ab}, Z$ in Eqs.~(\ref{Z_master_relation}) and $X_\rho$ 
in Eq.~(\ref{XHT_6dim_dynamical}) using Eq. (\ref{hypergeometric}) 
and obtain the expression of $H_L, H_T$, which we then insert in the first line.  Note that when evaluating $X_\rho$, one has to integrate the r.h.s. of the first line of Eq.~(\ref{XHT_6dim_dynamical}) and choose the integration constant so that the condition (\ref{H_tt_static_5D}) is satisfied, as mentioned before. Then, one finds in the leading order,
\begin{equation}
X_{(l)\rho} \sim -\frac{a_{0(l)}}{\rho} + O(1/\rho^3) . 
\end{equation}
Together with 
\begin{eqnarray}
Z_{(l)} \sim 4a_{0(l)} \frac{\rho^2}{L^2} + O(1) \,, \quad 
\end{eqnarray} 
one obtains  
\begin{equation}
H_L+ \frac{l(l+3)}{4}H_T = \frac{Z}{4}-2\rho (D^a \rho) X_a \sim 3a_{0(l)} \frac{\rho^2}{L^2} + O(1) \,,  
\end{equation}
which provides the expression of (\ref{g_chichi}). This $h_{\chi\chi}$ can be expanded as a series in $1/r$ by using Eq.~(\ref{chi_xi_relation}). The result is 
\be 
\lim_{\rho\to \infty} \displaystyle\frac{h_{\chi\chi}}{\rho^2} 
&=&\frac{5}{32}(48c_{(2)}-112c_{(4)}+189c_{(6)}) 
-\displaystyle\frac{3}{64L^2}(48a_{0(2)}-40a_{0(4)}+35a_{0(6)}) 
\nonumber \\
&+& \displaystyle\frac{5}{64r^2}\Bigl\{144a_{0(2)}-336a_{0(4)}+567a_{0(6)} 
 -16L^2(12c_{(2)}-154c_{(4)}+567c_{(6)})\Bigr\} 
\nonumber \\
&+&\displaystyle\frac{5}{32r^4}\Bigl\{-72a_{0(2)}+420a_{0(4)}-1323a_{0(6)} 
+L^2(96c_{(2)}-2576c_{(4)}+19089c_{(6)})\Bigr\}+O\left(\frac{1}{r^6}\right). 
\ee
From this expansion, the coefficients $\alpha_k$ in Eq.~(\ref{expansion_rr}) can be determined. 
Then, under the condition~(\ref{cond_alpha}), each coefficient $c_{(l)}$ can be obtained as 
\begin{align}
\label{c_coefficient}
& c_{(2)}=\frac{864a_{0(2)}+240a_{0(4)}+245a_{0(6)}}{5280L^2}, \nonumber \\
& c_{(4)}=\frac{-144a_{0(2)}+240a_{0(4)}+115a_{0(6)}}{2080L^2}, \nonumber \\
& c_{(6)}=-\frac{576a_{0(2)}+600a_{0(4)}-6895a_{0(6)}}{90090L^2}. 
\end{align}
Similarly, $h_L$ in Eq.~(\ref{def_h}) is expanded in terms of the harmonic functions $S_{(l)}$ as 
\begin{align}
\label{delta_h}
h_L=\sum_{l=2, 4, 6}\left(\frac{3a_{0(l)}}{L^2}\sin^2\chi  \mS_{(l)}
+c_{(l)}\sin\chi\cos\chi \mS_{(l), \chi}\right). 
\end{align}
Substituting Eq.~(\ref{chi_xi_relation}) into Eq.~(\ref{cond_h}), we determine each coefficient $a_{0(l)}~(l=2, 4, 6)$ in terms 
of the small parameter $M$ as 
\begin{align}
\label{parameter_a}
a_{0(2)}=-\frac{4M}{297}, \quad a_{0(4)}=-\frac{16M}{585}, \quad a_{0(6)}=-\frac{64M}{5005}. 
\end{align} 
Therefore the asymptotic bulk geometry is completely determined by the coefficients.  

Due to the gauge choice~(\ref{6dim_gauge_choice}), the coordinate transformation 
\begin{align}
\rho=\frac{L}{2}\left(\frac{1}{z}-z\right)
\end{align}
leads to the Fefferman-Graham coordinate~(\ref{Fefferman_Graham}) in the original 
conformally static Einstein frame, $x^\mu = (t, \chi, \theta^m)$, $m=1,2,3$. Then, under the boundary condition~(\ref{H_tt_static_5D}), 
${\tilde h}_{tt}$ in Eqs.~(\ref{6D_pert_variable}) can be expanded as a series in $z$,  
\begin{align}
\label{delta_g_tt_result}
& {\tilde h}_{tt}=
\sum_{l=2, 4, 6}\frac{4-l(l+3)}{L^2}\Biggl\{ \frac{3}{8}+\frac{3}{8}[1-l(l+3)] z^2
+\frac{4l(l+3)}{5} \left[ 
                            \frac{ \Gamma \left( \frac{3+l}{2} \right) }{ \Gamma \left( \frac{l+2}{2} \right) }
                     \right]^2 z^3
                                          \Biggr\}
a_{0(l)}\mS_{(l)}+O(z^4),  
\end{align} 
where we used Eq.~(\ref{a1_a0_relation}). So, according to the AdS/CFT dictionary, the energy density $T_{tt}$ 
in the static Einstein universe can be read off from the $z^3$ coefficient as 
\begin{align}
\label{negative_energy_5D}
 T_{tt} \simeq -\frac{3285 L^2 M}{1024}+O\left(\chi-\frac{\pi}{2}\right)^2 
\end{align}
combined with the result~(\ref{parameter_a}). Note that the Schwarzschild-AdS metric on the 
boundary~(\ref{asymptotic_D=5_Sch_AdS}) is obtained by a conformal transformation from 
the static Einstein frame by 
\begin{align}
\label{conformal_transformation}
g_{\mu\nu}\to e^{2\sigma}g_{\mu\nu}, \quad e^{\sigma}= \sqrt{f(r)} \,. 
\end{align}
Then, the stress-energy tensor is obtained from the transformation 
\begin{align}
T_{\mu\nu}\to e^{-3\sigma}T_{\mu\nu}. 
\end{align}
The asymptotic energy density in the $5$-dimensional Schwarzschild-AdS metric is obtained by 
\begin{align}
\label{asymptotic_energy_AdS}
T_{tt}\to {T_{tt}} f(r)^{-3/2} \sim r^{-3} \,. 
\end{align} 
Thus, the holographic energy~(\ref{holographic_energy}) becomes finite.    

Since the negative energy density~(\ref{negative_energy_5D}) in the asymptotic region is proportional 
to the mass parameter $M$ in the Schwarzschild AdS black hole~(\ref{D=5_Sch_AdS}), this reflects the 
curvature of the black hole. So, one may say that the negative energy density is caused by the vacuum 
polarization effect, as explained in~\cite{FT2013}. This behavior is qualitatively the same as the case of $5$-dimensional 
small black droplet numerical solutions~\cite{MarolfSantos2019} with an AdS black hole on the boundary.  

\section{Summary and discussions}
\label{sec:6}
In this paper, we have derived the first law of the entanglement entropy for two subsystems separated by an AdS black hole for 
odd-dimensional CFTs by using the holographic method and applying the Noether charge formula. We have seen that 
the nonvanishing contribution to our Noether charge generically arises not only at the bifurcate horizon and the asymptotic infinity but 
also at a spacelike hypersurface which terminates at the``equatorial plane" on the 
boundary static Einstein universe (see the dotted red line in Fig.~\ref{fig1}). This is because the boundary AdS black hole 
covers only one-half of the global AdS boundary.    

For $(2+1)$-dimensional CFT, we have, as its gravity dual, the exact black droplet solution~\cite{HubenyMarolfRangamani}, and therefore by perturbing it, we have holographically shown that the first law is satisfied without imposing any additional conditions. 
As for $(4+1)$-dimensional CFT, restricted to the $SO(4)$ symmetric perturbations, 
we have shown in Sec.~\ref{sec:4} that the first law is satisfied when there is no energy 
flux across the timelike conformal boundary, which corresponds to the Neumann boundary condition on the spacelike bulk hypersurface at $\chi=\pi/2$. 
We give a brief discussion on the analysis of generic perturbations in Appendix. We believe that an analysis for general perturbations without $SO(4)$ symmetry would also show the first law, but in the present paper, we have not been able to fully clarify this issue. 

We would like to emphasize again that in our holographic setup, in order to compute the Noether charge, we have taken $\Sigma$ as the 
lower half shaded region in Fig.~\ref{fig1} and imposed, in particular, the reflection (Neumann) boundary condition at $\chi=\pi/2$. 
The reflection boundary condition at $\chi=\pi/2$ is necessary to avoid additional contributions to the first law from the upper side of Fig.~\ref{fig1}. One may think of what happens if one takes $\Sigma$ as the entire time-slice. For example, if $\Sigma$ is taken as the union of the single black droplet in the lower half and the regular (no black droplet) region in the upper half of Fig.~\ref{fig1}, then the reflection condition at $\chi=\pi/2$ is no longer satisifed, and hence there is no guarantee that the first law can be derived. First of all, if $\Sigma$ is taken as the entire time-slice, then the corresponding boundary CFT would live in a spatially compact universe, rather than in an asymptotically AdS boundary spacetime, and the notion of a boundary black hole itself is no longer clearly defined. 

In Sec.~\ref{sec:5}, under the Neumann boundary condition for no-energy flux, we have constructed the asymptotic geometry of a small black droplet solution from scalar-type of the linear metric perturbations of the pure AdS spacetime. For this purpose, 
we have expanded the metric perturbations by scalar harmonics on the sphere 
and also set the boundary black hole mass $M$ as the small parameter. 
To satisfy the Neumann boundary condition, only even modes of the scalar harmonics are permitted. 

In Ref.~\cite{AvisIshamStorey1978}, quantization of conformally coupled scalar fields in anti-de Sitter spacetime 
was considered in various schemes. One is the ``transparent'' boundary condition in which each positive frequency 
classical solution propagates beyond the timelike boundary at spatial infinity, and hence the Klein-Gordon inner product 
is defined on the whole Einstein static cylinder, by adding another ``virtual'' AdS spacetime. In this case, 
observers living in an asymptotically AdS spacetime would recognize that some information or energy is lost through 
the timelike boundary. Another boundary condition is the ``reflective'' boundary condition in which each positive 
frequency classical solution is reflected at the timelike boundary at infinity, and therefore information or energy loss does not 
occur on the boundary. In this paper, by adapting the latter ``reflective'' boundary condition, we have shown the first law 
of the entanglement entropy for $D=5$ CFT in asymptotically AdS spacetime. Conversely, if the first law is satisfied in 
asymptotically AdS spacetime, the ``reflective'' boundary condition is derived for the holographic stress-energy tensor. 
On the other hand, in the $D=3$ case, the Noether charge at the infinity is always zero, being irrespective of 
the boundary conditions in the holographic model we consider. It would be interesting to explore whether the first 
law of the entanglement entropy is satisfied for various quantizations of a free scalar field in BTZ background.   

\section*{Acknowledgements}
We would like to thank Takashi~Okamura for useful discussions. This work was supported in part by 
JSPS KAKENHI Grant Numbers 17K05451, 20K03975~(KM) and 15K05092, 20K03938~(AI). 
 
\section{Appendix}
\label{sec:appendix}

In this appendix we consider the Noether charge formula (\ref{first_law_form}) for general perturbations in $(5+1)$-dimensional pure AdS bulk background metric~(\ref{global_AdS}). 
Note that although the perturbation analysis in Sec.~\ref{subsec:firstlaw} is restricted to $SO(4)$ symmetric case, the background geometry treated in Sec.~\ref{subsec:firstlaw} 
is the black droplet solution, which we assume to exist. In this sense the analysis in Sec.~\ref{subsec:firstlaw} is more relevant to our purpose for obtaining the first law than 
the analysis below.

As introduced in sec.~\ref{subsec:energyflux}, we use the coordinates $y^a := (t,\rho)$ in the $2$-dimensional 
AdS spacetime and the angular coordinates $x^i =( \chi, \theta^m)$. We denote our Killing vector field by $\xi^M=(\xi^a, \xi^i)$. 
The integrand of the Noether charge formula (\ref{first_law_form}) is in the present case a $4$-form 
on the $4$-sphere $(\Omega_{(4)}, \gamma_{ij})$. For generic metric perturbations $\delta g_{MN} = h_{MN}=(h_{ab}, h_{ai}, h_{ij})$, we have 
\bena\label{NC4form}
\delta {\bm Q}-\xi\cdot {\bm \Theta}(g,\, \delta g) 
&=& - \frac{1}{16\pi}\epsilon_{i_1\cdots i_4}{}^{ab}
\Bigg[ 
   \left(\frac{1}{2}h g^{ac} - h^{ac} \right)D_c\xi^b -h^{aj}D_j \xi^b + \frac{D^b \rho}{\rho}h^a_j \xi^j 
\nonumber \\
 &{}& \qquad \qquad \qquad 
 + \frac{1}{2} (D^ah^b_c-D^bh^a_c)\xi^c +\frac{\rho}{2}\left\{ D^a\left(\frac{h^b_j}{\rho}\right) - D^b\left(\frac{h^a_j}{\rho} \right) \right\} \xi^j
\nonumber \\
 &{}& \qquad \qquad \qquad - \xi^a\left\{
                          D_ch^{bc}+D_j h^{bj}+ n\frac{D_c\rho}{\rho}h^{bc}-\frac{D^b\rho}{\rho}h^k_k - D^b h
                 \right\}
\Bigg] .  
\eena 

The tensor-type metric perturbations $h_{MN}$ are tangential to $(\Omega_{(4)}, \gamma_{ij})$ with no components along $y^a$ direction, 
and behave as a transverse-traceless tensor on $(\Omega_{(4)}, \gamma_{ij})$. It is straightforward to see that the $4$-form (\ref{NC4form}) vanishes for tensor-type perturbations. 

Now let us consider the vector-type perturbations. We introduce the vector harmonics on $\Omega_{(4)}$, which satisfy 
\begin{align} 
\label{4dim_harmonic_vec}
D^iD_i {\mV}_{(l_v) j}+ [l_v(l_v+3)-1]\mV_{(l_v) j}=0, \quad D^i \mV_{(l_v) i}=0 , \quad l_v=1, 2, 3, \cdots .   
\end{align}
The vector-type metric perturbations can be expanded in terms of $\mV_{(l_v) i}$ as
\begin{align} 
\label{6D_pert_variable_vec}
& \delta g_{ab} = 0, \qquad \delta g_{ai}= \sum_{ l_v}  H^{(1)}_{(l_v)a} \mV_{(l_v)i} , 
\nonumber \\
& \delta g_{ij}= \sum_{l_v} H^{(1)}_{T(l_v)} \left(D_i {\mV}_{(l_v)j}  - D_j {\mV}_{(l_v)i} \right) .  
\end{align}
Hereafter we omit the mode indices $(l_v)$ for notational simplicity. Note that since the gauge transformation of the vector-type perturbations is generated by a vector field $X_i=X^{(1)}\mV_i$ with a scalar $X^{(1)}$ on the $2$-dimensional AdS spacetime, one can impose, for instance, 
the gauge condition $H^{(1)}_T=0$. 
   
The gauge invariant variables for the vector-type perturbations are given by 
\begin{align} 
\label{gauge-inv-variable_vec}
Z_a = H_a^{(1)} - \rho^2 D_a\left( \frac{H^{(1)}_T}{\rho^2} \right). 
\end{align}
This variable can be expressed by a single master variable $\Phi_V$ on the $2$-dimensional AdS spacetime spanned by $(t, \rho)$ as 
\begin{align} 
\label{gauge-inv-variable_vec}
Z_a = \frac{1}{\rho^2} \epsilon_{ab}D^b (\rho^2 \Phi_V) , 
\end{align}
where $\epsilon_{ab}$ is the metric compatible volume element on the $2$-dimensional AdS spacetime $g_{ab}$. The master variable $\Phi_V$ satisfies 
\begin{align} 
\label{master-eq_vec}
 D^aD_a \Phi_V - \left\{ \frac{2}{L^2} + \frac{2+l_v(l_v+3)}{\rho^2} \right\} \Phi_V =0. 
\end{align}
The normalizable solution $\Phi_V$ which also satisfies the regularity condition in the bulk is given by eq.~(154) of~~\cite{IshibashiWald2004} 
with $\nu = 3/2, \sigma=\sqrt{l_v(l_v+3)+9/4}$. (Note that for the normalizable solution, the frequency $\omega$ is quantized to be $\omega = \mp(2m +1+\nu+\sigma)$, 
$m=0,1,2, \dots$.)
Then, from eq.~(154) of~~\cite{IshibashiWald2004}, one can see the asymptotic behavior of $\Phi_V$ as 
\begin{align}
\Phi_V = \frac{b_{1}}{\rho^2}+\frac{b_{2}}{\rho^4} + \cdots.  
\end{align}
It then follows that at large $\rho$, 
\ben
\label{Ztrho}
Z_t \sim - \frac{2 b_2/L^2}{\rho^3} + O(1/ \rho^5) , \qquad Z_\rho \sim -i\omega L^2\left( \frac{b_1}{\rho^4} + \frac{b_2}{\rho^6} + \cdots \right) . 
\een

For the vector-type perturbations, the Noether charge formula~(\ref{NC4form}) is expressed in terms of the gauge-invariant variable 
(\ref{gauge-inv-variable_vec}) as
\bena
\label{NC4form_vec}
\delta {\bm Q}-\xi\cdot {\bm \Theta}(g,\, \delta g) 
= - \frac{1}{16\pi}\epsilon_{i_1\cdots i_4}{}^{ab}
\left[ 
   - \frac{Z_a}{\rho^2}\mV^iD_i \xi_b + D_{[a}Z_{b]} \xi^i \mV_i 
\right] . 
\eena 
So far, we have not yet used the property that $\xi^M$ is the Killing vector. In Sec.~~\ref{subsec:energyflux}, we are concerned with 
the static black hole background and our Killing vector is $\xi^a \partial_a=\partial/\partial t$, $\xi^i=0$. Thus, in this case, 
one can immediately see that the above formula (\ref{NC4form_vec}) vanishes. 

Now suppose that a rotating black droplet solution be available, and let us consider $\xi^i \neq 0$. In that case, one should be able to choose $\xi^i$ as a linear combination of the rotational symmetry generators of $\Omega_{(4)}$. 
It is known that for the dipole moment $l_v=1$, $\mV^i$ itself is a Killing vector field on $\Omega_{(4)}$, and therefore 
one can take $\xi^i$ one of such $\mV^i$ so that $\xi^i \mV_i \neq 0$. 
Then, one should, in principle, be able to find a Killing vector field $\xi^M \partial_M = \partial/\partial t + \Omega_{(A)} \partial/\partial \psi_{(A)}$ which becomes tangent to the null generators of the Killing horizon of the black droplet solution, where $\Omega_{(A)}$ are 
certain constants corresponding to the angular velocity of the black hole, and $\psi_{(A)}$ angular Killing parameters for $\xi^i$. 
For such a Killing vector field $\xi^M$, the second-term of r.h.s. of the above formula (\ref{NC4form_vec}) should provide the term proportional to the variation of the angular momenta. In fact, substituting (\ref{Ztrho}) into (\ref{NC4form_vec}) gives a nonvanishing term proportional to 
$\Omega_{(A)}(\omega^2 L^2 b_1 + 6b_2/L^2)$. For the stationary case $\omega =0$, this can be viewed as adding an angular momentum to the background black hole by the dipole perturbation.  

The analysis of generic perturbations of the scalar-type appears to be much more involved with more terms in the Noether charge formula (\ref{NC4form}). If we can evaluate the integration of the above formula (\ref{NC4form}) for both the vector-type and scalar-type 
perturbations, not only at $\rho \rightarrow \infty, \, \chi \rightarrow \pi/2$ but also at the Killing horizon of the (hypothetical) rotating black droplet as well as other relevant boundaries, then we expect that the first law with work term should be obtained. 
To fully verify this argument is, however, beyond the scope of the present paper where we consider only linear perturbations and their asymptotic behavior.


\end{document}